\documentclass[twocolumn]{emulateapj}

\slugcomment{To be published in The Astronomical Journal}

\shorttitle{Signposts of Multiple Bipolar Ejections in NGC\,6881}
\shortauthors{Ramos-Larios, Guerrero, \& Miranda}

\begin{document}

\title{THE UNUSUAL DISTRIBUTIONS OF IONIZED MATERIAL AND MOLECULAR HYDROGEN 
       IN NGC\,6881: SIGNPOSTS OF MULTIPLE EVENTS OF BIPOLAR EJECTION 
       IN A PLANETARY NEBULA}

\author{Ramos-Larios, G.\altaffilmark{1}}
\affil{
Instituto de Astronom\'{\i}a y Meteorolog\'{\i}a, Av. Vallarta No. 2602, 
Col.\ Arcos Vallarta, C.P. 44130 Guadalajara, Jalisco, M\'exico; 
gerardo@astro.iam.udg.mx}

\author{Guerrero, M.A.\ and Miranda, L.F.}

\affil{
Instituto de Astrof\'{\i}sica de Andaluc\'{\i}a, IAA-CSIC, 
C/ Camino Bajo de Hu\'etor 50, 18008 Granada, Spain; 
mar@iaa.es, lfm@iaa.es}

\altaffiltext{1}{
Visiting Astronomer, Instituto de Astrof\'{\i}sica de Andaluc\'{\i}a. }

\begin{abstract}

The planetary nebula NGC\,6881 displays in the optical a quadrupolar 
morphology consisting of two pairs of highly collimated bipolar lobes 
aligned along different directions.  
An additional bipolar ejection is revealed by the hydrogen molecular 
emission, but its wide hourglass morphology is very different from 
that of the ionized material.  
To investigate in detail the spatial distribution of molecular hydrogen 
and ionized material within NGC\,6881, and to determine the prevalent 
excitation mechanism of the H$_2$ emission, we have obtained new near-IR 
Br$\gamma$ and H$_2$ and optical H$\alpha$ and [N~{\sc ii}] images, as 
well as intermediate resolution ${JHK}$ spectra.
These observations confirm the association of the H$_2$ bipolar lobes to 
NGC\,6881 and find that the prevalent excitation mechanism is collisional.  
The detailed morphology and very different collimation degree of 
the H$_2$ and ionized bipolar lobes of NGC\,6881 not only imply 
that multiple bipolar ejections have occurred in this nebula, but 
also that the dominant shaping agent is different for each bipolar 
ejection:  a bipolar stellar wind most likely produced the H$_2$ lobes, 
while highly collimated outflows are carving out the ionized lobes into 
the thick circumstellar envelope.  
The asymmetry between the southeast and northwest H$_2$ bipolar 
lobes suggests the interaction of the nebula with an inhomogeneous 
interstellar medium.  
We find evidence that places NGC\,6881 in the H~{\sc ii} region Sh\,2-109 
along the Orion local spiral arm.

\end{abstract}

\keywords{
infrared: ISM: continuum --- 
ISM: molecules --- 
planetary nebula: individual (NGC 6881)}

\section{Introduction}

Asymmetry is common among planetary nebulae (PNe) and it comes in 
a large variety of shapes and morphological features.  
On a large scale, the departure from symmetry goes from the mild 
asymmetry of elliptical PNe to the strong asymmetry displayed by 
bipolar PNe.  
Asymmetry in PNe is also revealed by an assortment of small-scale 
morphological features including the symmetric fast low-ionization emission regions (FLIERs) of elliptical PNe, as well as point-symmetric collimated outflows and their 
associated blowouts and bow-shock structures.  
The asymmetry in PNe has been linked on many occasions to binarity 
\citep[e.g.,][]{S06}, but there is no conclusive evidence supporting 
such connection \citep[][and references therein]{SP04,B00,S97}.  
On the contrary, the onset of bipolarity in PNe seems to require 
the evolution of a massive progenitor \citep[e.g.,][]{CS95}.

PNe that show asymmetric structures along different directions are of great 
importance, as the change in the direction of the ejection can be associated 
with the precession of the progenitor star in a binary system.  
This class of PNe is not homogeneous at all, and includes PNe with multiple 
collimated outflows along different directions \citep{Cetal97} or 
with precessing jet-like features \citep{MGT99}, 
starfish-shaped and multipolar PNe that show point-symmetric bow-shock 
features \citep{ST98,S00}, and quadrupolar and polypolar 
PNe with different sets of bipolar lobes aligned along different symmetry 
axes \citep{MSG96,Letal98}.  
Polypolar and quadrupolar PNe are especially interesting as multiple 
hourglass structures imply recurrent bipolar ejections which are 
particularly difficult to interpret in the framework of the interacting 
stellar wind model \citep{KPF78,B87}.

One of the most intriguing PNe with multiple bipolar lobes is NGC\,6881.  
Originally classified as a quadrupolar PN, based on optical narrowband 
images and long-slit echelle spectra \citep{GM98}, NGC\,6881 presents 
two pairs of highly collimated bipolar lobes with very similar, although 
not coincident, symmetry axes.  
The southwest lobe displays a loop-like feature that is highly 
reminiscent of a precessing collimated outflow, but its measured 
expansion velocity is low.  
\citet{KS05} showed that the dense equatorial ring has recently changed 
its orientation, being aligned with the youngest pair of bipolar lobes.

Therefore, there are many signs suggesting that precession and sequential 
events of bipolar ejections have occurred in NGC\,6881.  
The spatial distribution of the molecular hydrogen in this nebula, 
as revealed by near-IR narrowband H$_2$ (1--0) S(1) images \citep{Getal00}, 
adds a new twist.  
Molecular hydrogen emission is detected in wide hourglass bipolar 
lobes that extend much farther than the ionized bipolar lobes.  
The H$_2$ bipolar lobes may represent a bipolar ejection unrelated 
to the formation of the two pairs of ionized bipolar lobes.

To study carefully the spatial distributions of ionized 
material and molecular hydrogen in NGC\,6881 and to investigate the 
excitation mechanism of the H$_2$ molecule throughout the nebula, 
we have obtained new H$_2$, Br$\gamma$, [N~{\sc ii}], and H$\alpha$ 
images, and $JHK$ intermediate-resolution long-slit spectroscopic 
observations.  
A description of the observations is presented in $\S$2 
and the results are given in $\S$3.  
The results are discussed in $\S$4, and the conclusions and a short 
summary are presented in $\S$5.

\begin{table}[!t]\centering
\setlength{\columnwidth}{0.3\columnwidth}
\setlength{\tabcolsep}{2.5\tabcolsep}
\caption{Narrow-Band Imaging of NGC 6881}
\begin{tabular}{llrrr}
\hline
\multicolumn{1}{l}{Telescope} & 
\multicolumn{1}{l}{Filter} & 
\multicolumn{1}{c}{$\lambda$$_{c}$} & 
\multicolumn{1}{c}{$\Delta\lambda$} & 
\multicolumn{1}{c}{Exp.\ Time} \\
\multicolumn{1}{c}{} & 
\multicolumn{1}{c}{} & 
\multicolumn{1}{c}{(\AA)} & 
\multicolumn{1}{c}{(\AA)} & 
\multicolumn{1}{c}{(s)} \\ 
\hline
WHT &  H$_2$(1$-$0)  & 21218 & 320 &  900~~~~ \\
WHT &  Br$\gamma$    & 21658 & 320 & 1000~~~~ \\
WHT &  Kc            & 22700 & 340 & 1000~~~~ \\
\hline				  			       
NOT &  [N~{\sc ii}]  &  6589 &   9 & 1800~~~~ \\
NOT &  H$\alpha$     &  6568 &   8 &  900~~~~ \\
    \hline
  \end{tabular}
\vspace{0.4cm}
\end{table}

\section{Observations}

\subsection{Imaging} \label{im}

Narrowband near-IR images of NGC\,6881 were obtained during 2006 
September 9 using LIRIS (Long-Slit Intermediate Resolution Infrared 
Spectrograph) at the Cassegrain focus of the 4.2 m William Herschel 
Telescope (WHT) on Roque de Los Muchachos Observatory (ORM, 
La Palma, Spain).  
The detector was a 1k$\times$1k HAWAII array sensitive in the spectral 
range from 0.8 to 2.5 $\mu$m.  
The plate scale is 0$\farcs$25 pixel$^{-1}$ and the field of view (FOV) 
is 4$\farcm$27$\times$4$\farcm$27.  
The narrowband filters isolated the H${_2}$~(1$-$0)~S(1) 2.1218 $\mu$m 
and Br$\gamma$ 2.1658 $\mu$m emission lines.  
An additional narrowband filter (Kc) centered at 2.270 $\mu$m was 
used in order to subtract the continuum contribution from the line 
emission detected through the H${_2}$ and Br$\gamma$ filters.  
The central wavelength and bandwidth of these filters are listed in Table~1. 

\begin{figure*}[!t]
\includegraphics[width=\columnwidth]{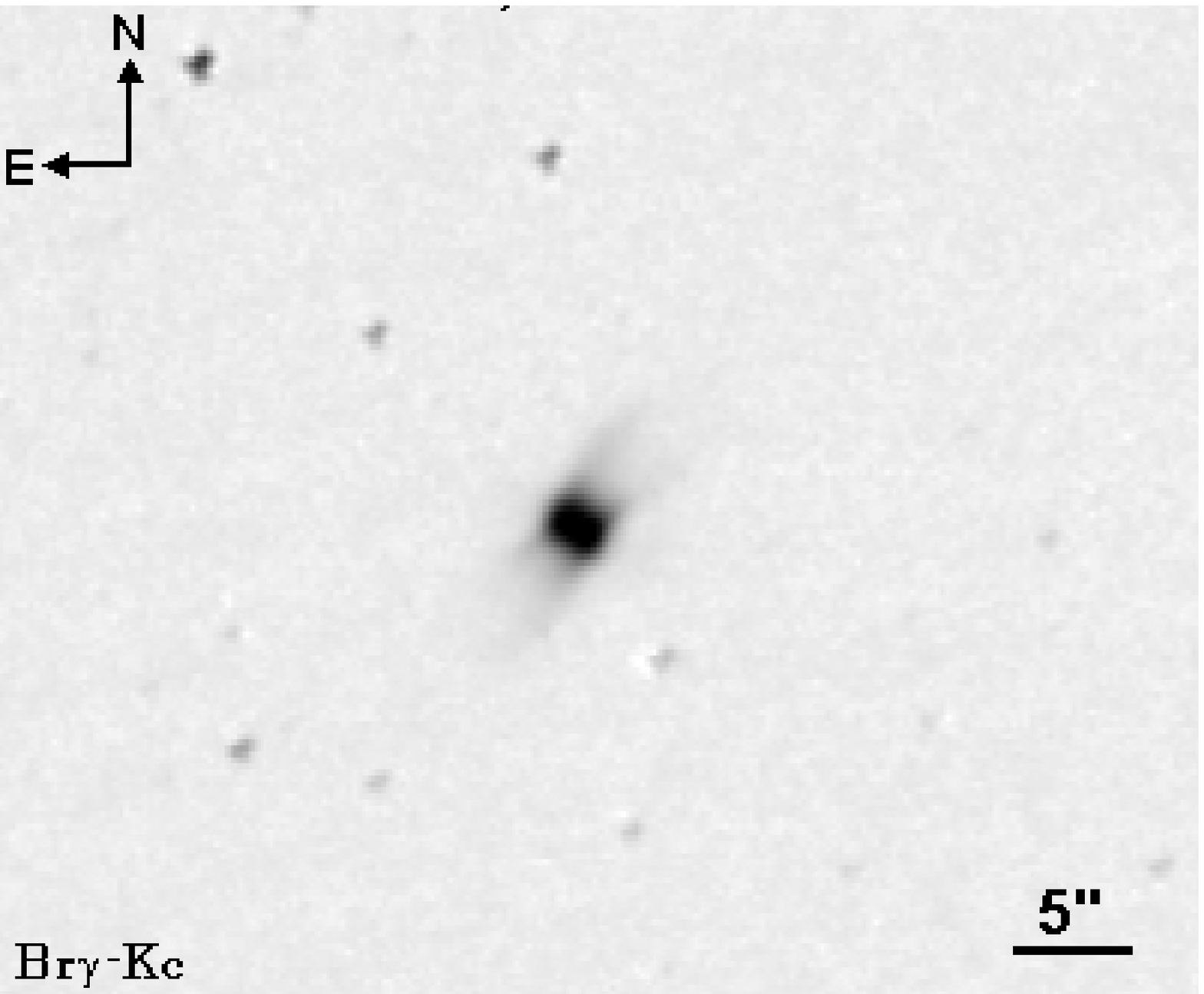}%
\hspace*{\columnsep}%
\includegraphics[width=\columnwidth]{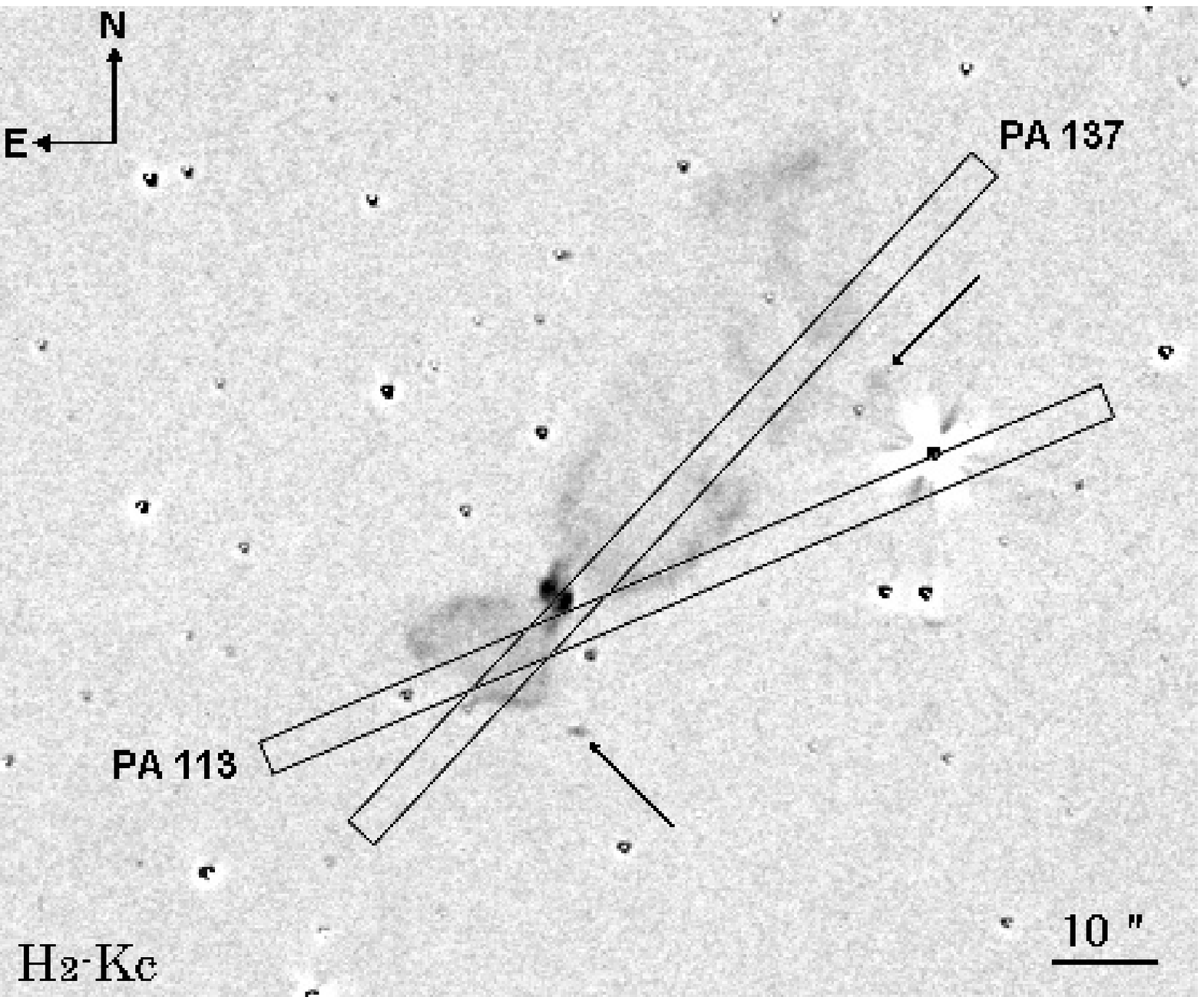}
\vskip .1in
\includegraphics[width=\columnwidth]{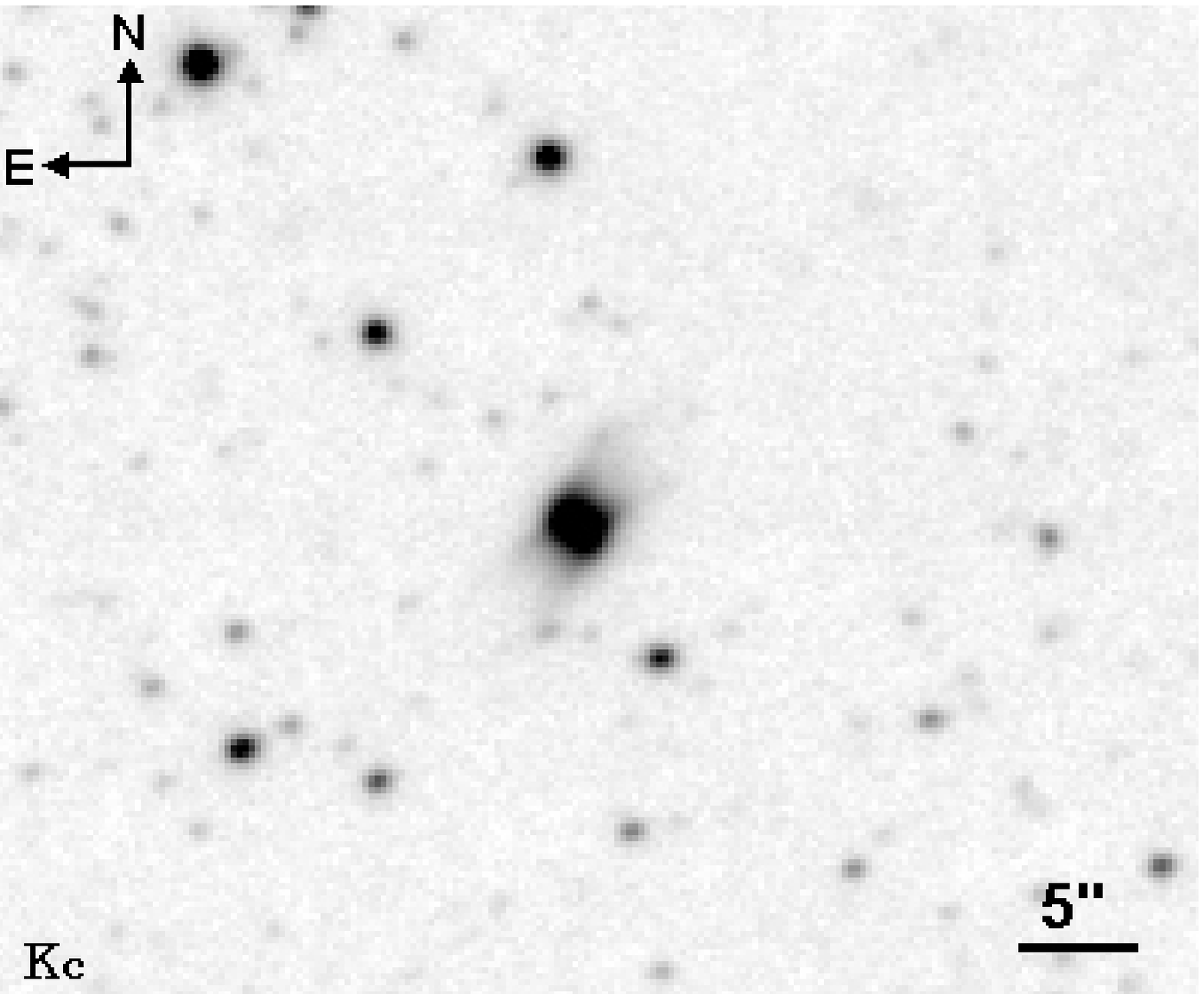}%
\hspace*{\columnsep}%
\includegraphics[width=\columnwidth]{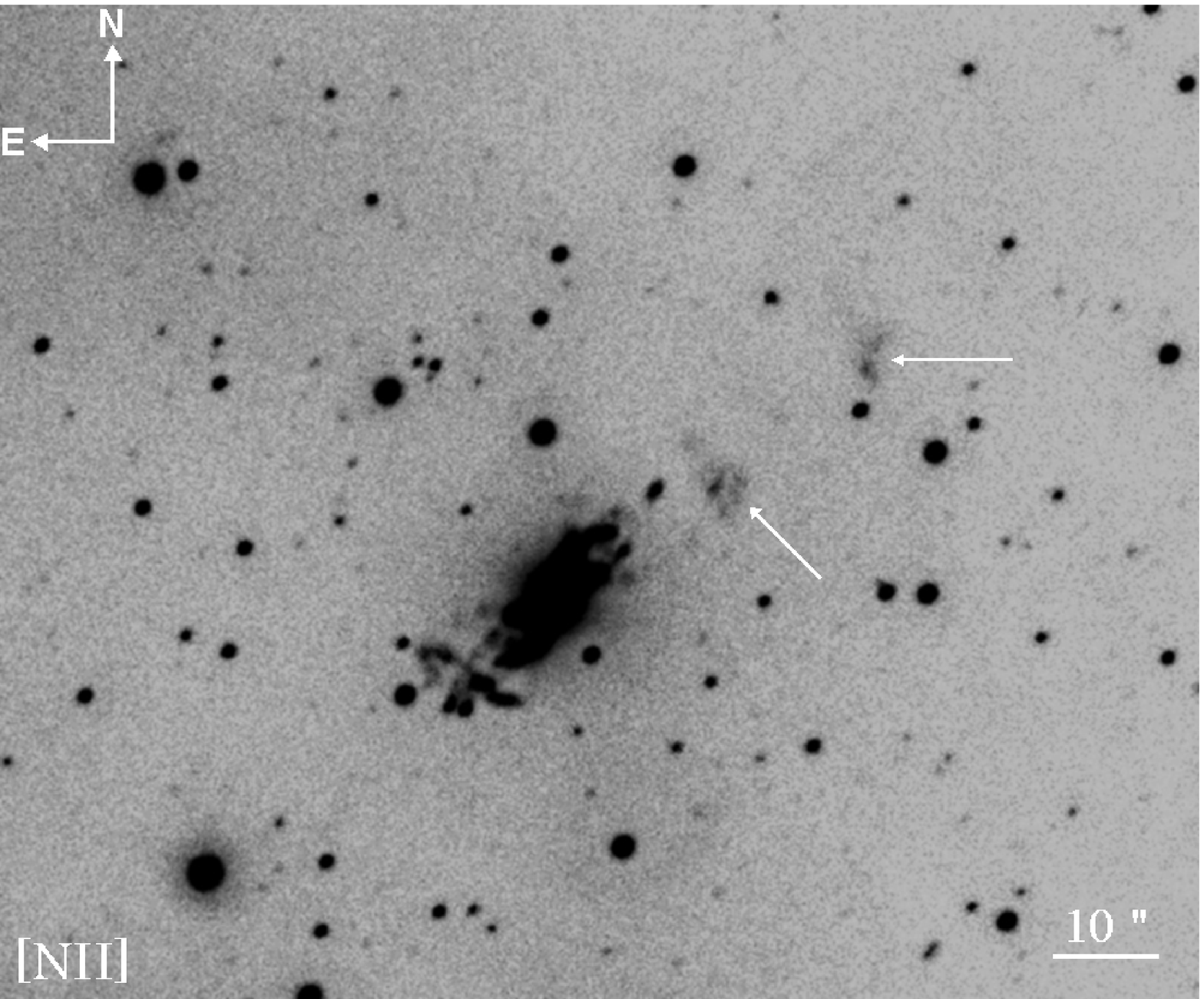}
\vskip .1in  
\includegraphics[width=\columnwidth]{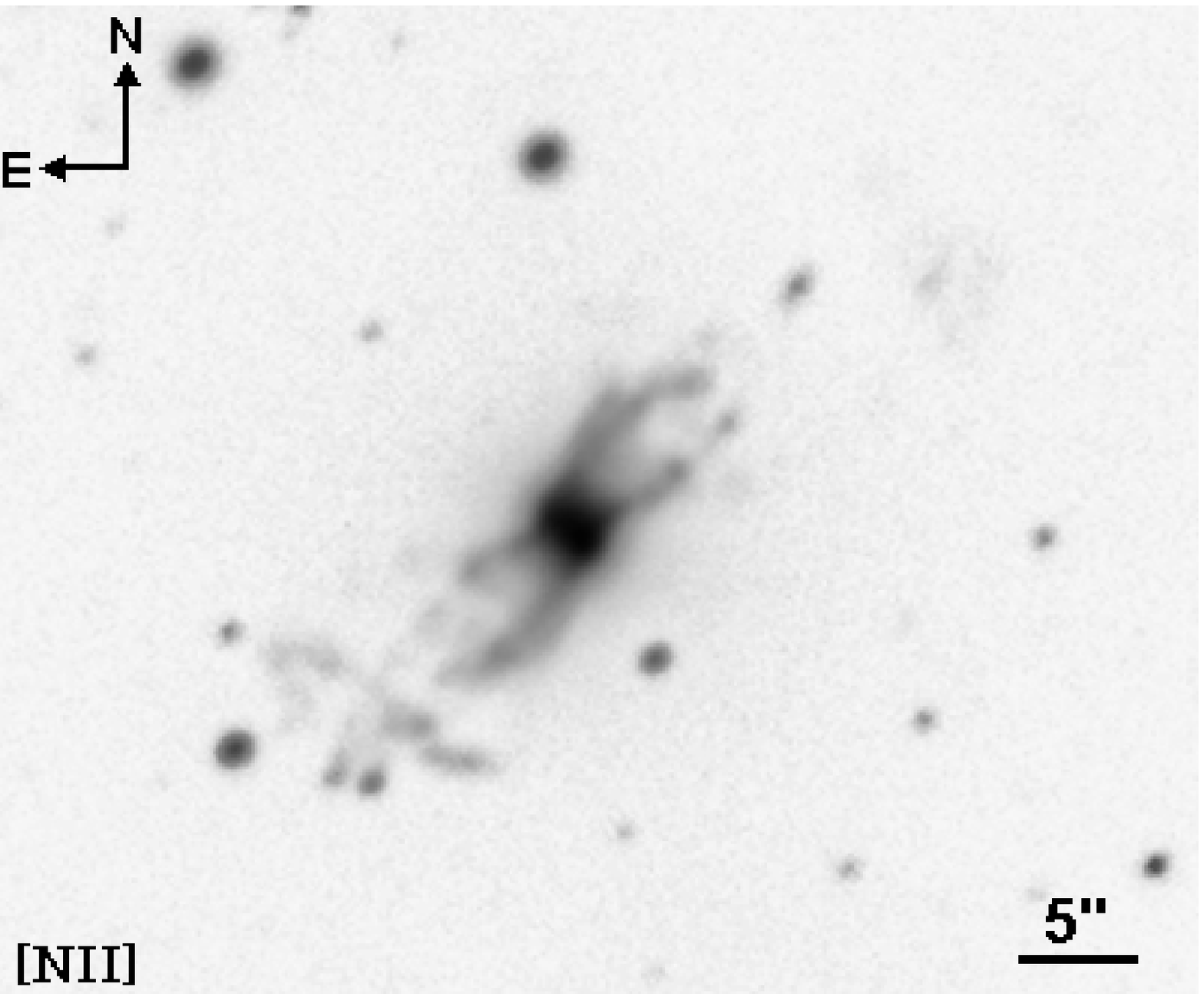}
\hspace*{\columnsep}%
\includegraphics[width=\columnwidth]{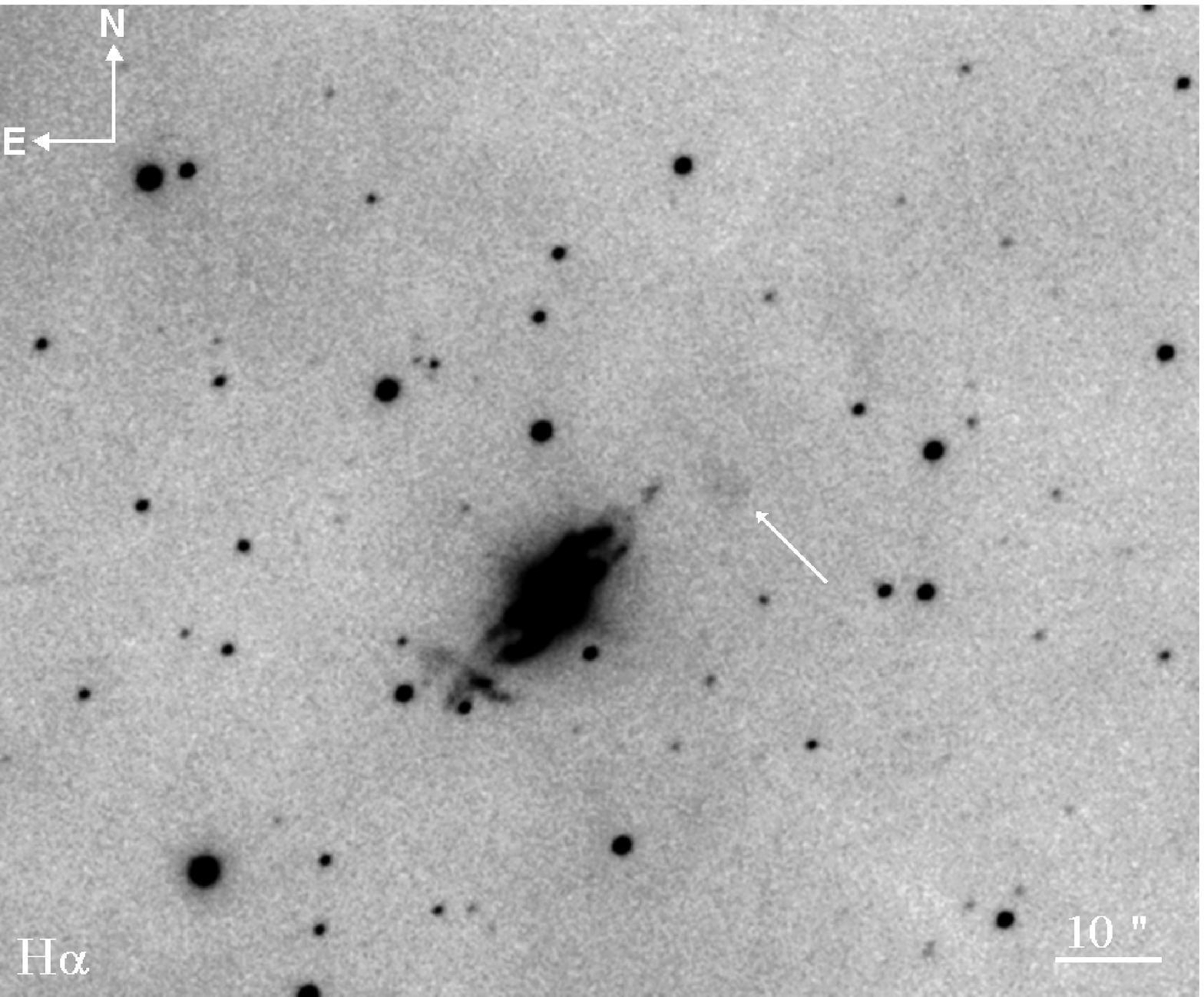}%
\caption{
Images of NGC\,6881 in the Br$\gamma$ {\it (top, left)}, Kc 
{\it (middle, left)}, and [N~{\sc ii}] {\it (bottom, left)} 
narrowband filters.  
The images in the H$_2$ {\it (top, right)}, [N~{\sc ii}] {\it (middle, right)} 
and H$\alpha$ {\it (bottom, right)} filters show a larger field of view than 
the images in the left column.  
The artifacts noticeable around the field stars in the continuum 
subtracted Br$\gamma$ and H$_2$ images are residuals of the 
subtraction caused by a slight mismatch of the PSF between the line 
and continuum images.  
The arrows towards the northwest of the nebula in the right panels 
mark diffuse emission in the H$_2$, [N~{\sc ii}], and H$\alpha$ 
images associated with the northwest H$_2$ lobe and its northwest 
extension.
The arrow towards the south of the nebula in the H$_2$ image 
marks diffuse emission unrelated to NGC\,6881, probably a 
background galaxy.  \\
}
\label{fig1}
\end{figure*}

We obtained series of 100 s exposures on each filter for the total 
integration times given in Tab.~1.  
The telescope pointing was shifted by a few pixels between each exposure, 
rastering the nebula to different locations on the detector.  
Each series of observations on the object was followed by a similar series 
of observations on adjacent blank sky positions.
Individual exposures were flat-fielded and dark-corrected, and the sky 
frames were combined to obtain a master sky image.  
The images on the nebula were subsequently sky-subtracted, shifted, and 
averaged.  
The continuum contribution to the H${_2}$ and Br$\gamma$ emission was 
removed using the continuum Kc image.  
Since the H${_2}$, Br$\gamma$, and Kc filters have very similar bandwidth, 
we do not find it necessary to scale the images, although differences in the point-spread function (PSF) of stars in the field of view required us to degrade the PSF of the H${_2}$ image to match these of the Br$\gamma$ and Kc images.  
The continuum subtracted H${_2}$ and Br$\gamma$ images and the Kc image 
are presented in Figure~\ref{fig1}.  
The spatial resolution, as determined from the FWHM of stars 
in the FOV, is $\sim$0$\farcs$8.

Narrowband images in the [N~{\sc ii}] and H$\alpha$ emission lines were 
obtained on 2006 June 30 using ALFOSC (Andalucia Faint Object 
Spectrograph and Camera) at the 2.56 m Nordic Optical Telescope (NOT) 
in the ORM.  
The camera is a 2048$\times$2048 CCD with a plate scale of 
0$\farcs$19~pixel$^{-1}$ and a FOV of 6$\farcm$5$\times$6$\farcm$5.  
The [N~{\sc ii}] and H$\alpha$ images of NGC\,6881 are also presented 
in Fig.~\ref{fig1}.  
The spatial resolution, as determined from the FWHM of stars in the 
FOV, was 0$\farcs$9.

\begin{table}[!t]\centering
\setlength{\columnwidth}{0.001\columnwidth}
\setlength{\tabcolsep}{1.2\tabcolsep}
\caption{NIR Spectroscopic Observations of NGC\,6881}
\begin{tabular}{lcccrcc}
\hline
\multicolumn{1}{c}{Grism} & 
\multicolumn{1}{c}{$\lambda$} & 
\multicolumn{1}{c}{Dispersion} & 
\multicolumn{1}{c}{Slit Width} & 
\multicolumn{1}{c}{~~R} & 
\multicolumn{1}{c}{PA} & 
\multicolumn{1}{c}{Exp time} \\ 
\multicolumn{1}{c}{} & 
\multicolumn{1}{c}{($\mu$m)} & 
\multicolumn{1}{c}{(\AA~pix$^{-1}$)} & 
\multicolumn{1}{c}{($\arcsec$)} & 
\multicolumn{1}{c}{} & 
\multicolumn{1}{c}{($\arcdeg$)} & 
\multicolumn{1}{c}{(s)} \\ 
\hline
$JH$        & 1.15$-$1.75 & 6.6 & 0.75 &  $\sim$575 & 113 & 1800 \\
$K_{\rm B}$ & 1.92$-$2.34 & 4.3 & 0.75 & $\sim$1700 & 113 & 3000 \\
$K_{\rm B}$ & 1.92$-$2.34 & 4.3 & 0.75 & $\sim$1700 & 137 & 1200 \\
\hline
\end{tabular}
\vspace{0.4cm}
\end{table}

\subsection{Spectroscopy}

Intermediate-resolution $JHK$ long-slit spectroscopic observations 
were performed on 2004 July 2-4 using the NICS (Near Infrared Camera 
Spectrometer) at the 3.5 m Telescopio Nazionale Galileo (TNG) 
at the ORM.  
The NICS is a multimode instrument for IR observations (0.9$-$2.5 $\mu$m) 
that uses a HAWAII $1024\times 1024$ array as detector.  
We used the LF camera, providing a plate scale of 0$\farcs$25~pixel$^{-1}$, 
and the 0.75 slit, with a width of 0$\farcs$75 and a length of 4$'$.  
Observations were obtained using the $JH$ and $K_{\rm B}$ grisms 
and the long-slit was placed along the central star at P.A.s 113$\arcdeg$ 
and 137$\arcdeg$ (Table~2).  
In order to provide the means for subtracting the sky emission, 
the nebula was placed at different positions along the slit.

The spectra were reduced using IRAF\footnote{
IRAF is distributed by the National Optical Astronomy Observatory 
which is operated by the Association of Universities for Research in
 Astronomy, Inc. under cooperative agreement with the National 
Science Foundation}
routines of the noao.twodspec and noao.onedspec packages.  
The data were flat-fielded using dome flats, and the sky contribution 
was removed using sky spectra obtained at the same time as the nebular 
spectra. 
We used telluric sky lines for the wavelength calibration.
The standard IR stars of spectral type A$-$A0 SAO\,15832, SAO\,48300, 
and SAO\,72320 \citep{Hetal98} were used for the flux calibration.  
We obtained the sensitivity function by comparing the spectra of these 
stars with a blackbody model of temperature 9480 K, similar to the 
effective temperature of A0 stars.  
Finally, the telluric absorptions in the one-dimensional spectra of 
NGC\,6881 were removed using the IRAF task ``telluric'', applying 
it to stars with featureless spectra in the $JHK$ range.

\section{Results}

\subsection{Morphology}

\begin{figure}[!t]
\begin{center}
\includegraphics[width=0.99\linewidth]{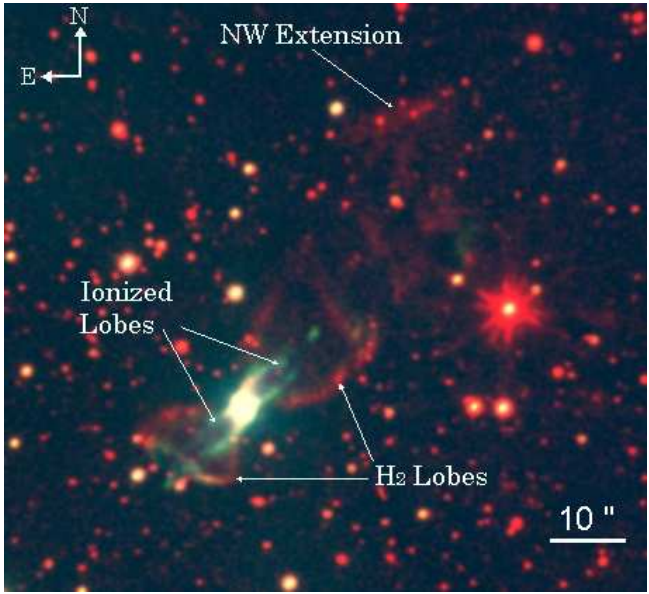}
\caption{ WHT LIRIS H$_2$ (red), and NOT ALFOSC H$\alpha$ (blue) and 
[N~{\sc ii}] (green) composite picture of NGC\,6881 overlaid with the 
different morphological features identified in the nebula. 
}
\label{fig2}
\end{center}
\end{figure}

The notable differences in the spatial distributions of ionized 
material and molecular hydrogen within NGC\,6881 described by 
\citet{Getal00} are specially highlighted in the H$_2$, H$\alpha$, 
and [N~{\sc ii}] composite picture shown in Figure~\ref{fig2}.  
In NGC\,6881, we can distinguish the following components.

\smallskip
\subsubsection  {The Equatorial Torus at the Central Region}

\begin{figure}[!t]
\begin{center}
\fbox{\includegraphics[width=0.88\columnwidth]{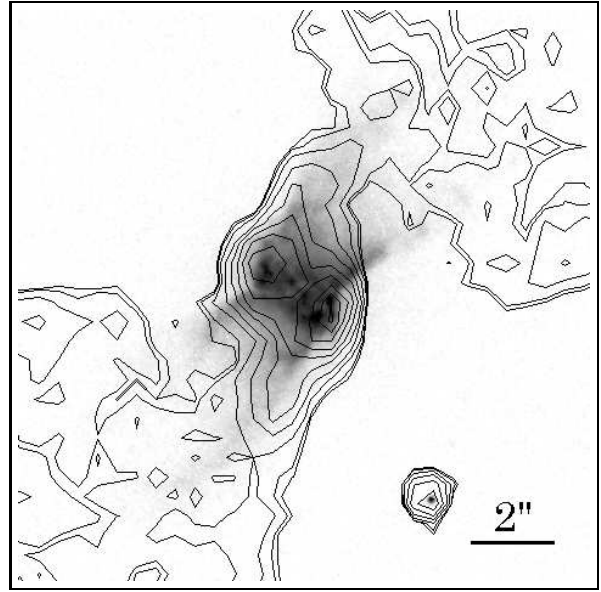}}
\caption{
Expanded view of the central region of NGC\,6881 in the \emph{HST} WFPC2 
[N~{\sc ii}] image.  
The image is overlaid with H$_2$ contours extracted from the WHT 
LIRIS continuum-subtracted H$_2$ image.  
} 
\label{fig3}
\end{center}
\end{figure}

The central regions of NGC\,6881 contain both molecular and ionized 
material.  
This region can be described as a clumpy torus-like structure \citep{KS05} 
that is expanding at a moderate velocity \citep{GM98}.  
The details of the molecular and ionized material distribution in the 
central regions of NGC\,6881, however, could not be studied properly 
by \citet{Getal00}, because of the limited spatial resolution of their 
near-IR images and poor continuum subtraction at this region.
Our new continuum-subtracted H$_2$ image reveals two bright knots of H$_2$ 
emission located at the central region of NGC\,6881 along its minor axis, 
suggesting a ring of molecular material.  
The relative distribution of ionized and molecular material in the 
central region of NGC\,6881, shown in Figure~\ref{fig3}, reveals 
that the H$_2$ emission encompasses the [N~{\sc ii}] emission.  
 
Molecular hydrogen in this region survives in a thin layer surrounding 
the ionized torus-like structure, most likely shielded by dense material 
from the ionizing flux of the central star.

\smallskip
\subsubsection {The Highly Collimated Bipolar Lobes}

The two pairs of highly collimated bipolar lobes are mainly detected 
in the Br$\gamma$, [N~{\sc ii}], and H$\alpha$ images, while the H$_2$ 
emission from these lobes is weak.  
The northern wall of the northwest lobe is brighter than its southern
wall, and the opposite applies to the southeast lobe, thus showing the 
point-symmetric brightness distribution typical of other bipolar PNe 
\citep[e.g., K\,4-55,][]{GMS96}.

\smallskip
\subsubsection  {The Hourglass H$_2$ Bipolar Lobes}

Contrary to the highly collimated bipolar lobes, the open hourglass bipolar 
lobes of NGC\,6881 are dominated by H$_2$ emission.  
  
They show a narrow waist coincident with NGC\,6881 central regions 
and clear limb-brightening, indicating that they are formed by a 
thin layer of material.  

\begin{figure}[!t]
\begin{center}
\includegraphics[width=0.9\columnwidth]{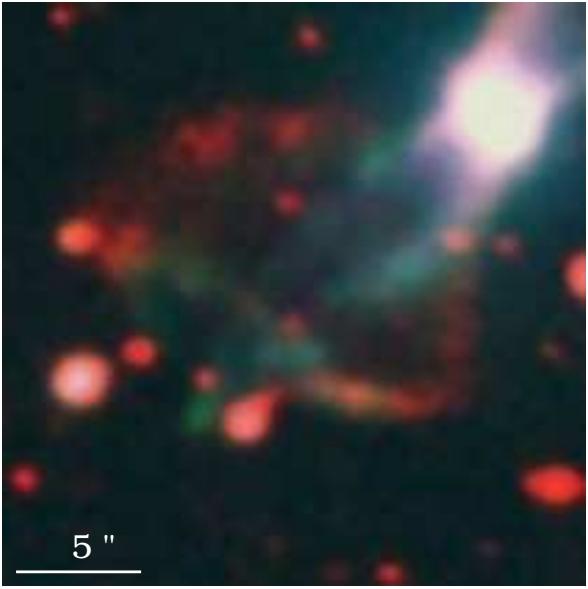}
\caption{
Expanded view of the southeast loop-like feature from the H$_2$, 
H$\alpha$, and [N~{\sc ii}] composite picture of NGC\,6881 (Fig.~\ref{fig2}).  
}
\label{fig4}
\end{center}
\end{figure}

The southeast lobe, extending $\sim$12\arcsec, is smaller than 
the northwest lobe, with a size up to $\sim$20\arcsec.  
There are notable differences between the tips of the southeast 
and northwest lobes.  
The tip of the southeast lobe is sharp and can be associated 
with the southeast loop-like structure detected in H$\alpha$ 
and very prominently in [N~{\sc ii}] (Figure~\ref{fig4}).  
The westernmost emission of the [N~{\sc ii}] loop is coincident with the 
southwestern elbow of the H$_2$ lobe, and the [N~{\sc ii}] loop structure 
traces closely the edge of the H$_2$ lobe.  
The northwest lobe does not have a sharp edge nor does it displays 
a loop-like structure, but only a knot and a thin filament 
visible in the [N~{\sc ii}] image located at its Westernmost tip 
(Figs.~\ref{fig1} and \ref{fig2}).

\smallskip
\subsubsection  {The Northwest Bipolar Lobe Extension}

The H$_2$ emission of the northwest hourglass lobe displays an extension 
that casts its emission outwards up to $\sim$50\arcsec\ from the center 
of NGC\,6881 along its symmetry axis.  
This emission is mainly detected in H$_2$, but there are also
hints of [N~{\sc ii}] and H$\alpha$ emission (Fig.~\ref{fig2}).  
The H$_2$ surface brightness of this feature is not homogeneous, but 
is distributed on a series of three bright bands interspersed between bands 
of diminished brightness.  
Interestingly, the bright bands can also be recognized in the 
H$\alpha$ image.  
The northwesternmost tip of these bands shows a remarkable wedge-shaped 
morphology.

\begin{table}[!h]\centering
 \setlength{\tabcolsep}{1.1\tabcolsep}
\caption{ Line Fluxes at P.A. 137$^{\circ}$ with $K_{\rm B}$ Grism}
\begin{tabular}{llccccc}
\hline
\multicolumn{2}{c}{} & \multicolumn{4}{c}{Regions} \\
\cline{3-6}
\multicolumn{1}{c}{$\lambda$} & \multicolumn{1}{c}{Line ID} & ~Central~ & ~Ionized~ & H$_2$ & NW H$_2$ lobe  \\
\multicolumn{1}{c}{} & \multicolumn{1}{c}{} & \multicolumn{1}{c}{Regions} & lobes & lobes & extension \\
\multicolumn{1}{c}{[\AA]} & \multicolumn{1}{c}{} & 
\multicolumn{4}{c}{[$\times$10$^{15}$~ergs~~cm$^{-2}$~~s$^{-1}$~~\AA$^{-1}$~]} \\
\hline
19446 &	Br8 H~{\sc i}  & 63  & $\dots$ & $\dots$ & $\dots$ \\
19549 &	He ~{\sc i}	   & 16  & $\dots$ & $\dots$ & $\dots$ \\
19574 &	H$_2$ (1,0) S(3) & 12  & 17      & 11      & 9.2     \\
19703 &	H$_2$ (8,6) O(2) & 2.1 & $\dots$ & $\dots$ & $\dots$ \\
20338 &	H$_2$ (1,0) S(2) & 4.2 & 2.1     & 3.8     & 4.6     \\
20377 &	?	         & 4.7 & $\dots$ & $\dots$ & $\dots$ \\
20587 &	He ~{\sc i}  & 47  & 1.3     & 0.4     & 2.4     \\
20732 &	H$_2$ (2,1) S(3) & 1.2 & 0.8     & 1.3     & 2.8     \\
20763 &	?	         & 0.9 & $\dots$ & $\dots$ & $\dots$ \\
21126 &	He ~{\sc i}	  & 5.1 & $\dots$ & $\dots$ & $\dots$ \\
21138 &	He ~{\sc i}	  & 2.2 & $\dots$ & $\dots$ & $\dots$ \\
21218 &	H$_2$ (1,0) S(1) & 12  & 4.7     & 5.6     & 6.2     \\
21542 &	H$_2$ (2,1) S(2) & 0.4 & $\dots$ & $\dots$ & $\dots$ \\
21614 &	He ~{\sc i} ?    & 2.9 & $\dots$ & $\dots$ & $\dots$ \\
21658 &	Br7 H~{\sc i} (Brg)	 & 110 & 5.5     & 1.3     & $\dots$ \\
21793 &	Br ~{\sc i}	  & 1.1 & $\dots$ & $\dots$ & $\dots$ \\
21887 &	He ~{\sc ii}	  & 17  & 0.8     & $\dots$ & $\dots$ \\
21985 &	?	         & 2.1 & $\dots$ & $\dots$ & $\dots$ \\
22235 &	H$_2$ (1,0) S(0) & 3.3 & 1.2     & $\dots$ & 1.5     \\
22477 &	H$_2$ (2,1) S(1) & 1.6 & $\dots$ & $\dots$ & 1.0     \\
22872 &	Br ~{\sc i} ?    & 7.1 & $\dots$ & $\dots$ & $\dots$ \\
\hline
\end{tabular}
\vspace{0.4cm}
\end{table}

\smallskip
\subsubsection  {Large-Scale Emission}

The H$\alpha$ and [N~{\sc ii}] images of NGC\,6881 show large-scale 
diffuse emission, with the nebula laying on a broad arc of patchy 
emission that crosses the FOV from the northeast to the south 
(Fig.~\ref{fig1}).  
An examination of 
an H$\alpha$ image from the INT/WFC Photometric H$\alpha$ Survey (IPHAS) 
of the northern Galactic plane \citep{Detal05} of this region reveals a 
complex system of filaments.  
NGC\,6881 ($l$=74\fdg5520, $b$=+02\fdg1137) is projected along the 
local spiral arm onto the intricate network of H$\alpha$ filaments 
near the proximity of $\gamma$ Cyg \citep[e.g.,][]{PGK79}.  
In particular, NGC\,6881 is projected towards the east of a large cavity of 
size 20\arcmin$\times$12\arcmin\ oriented along the north-south direction 
that forms part of the H~{\sc ii} region Sh\,2-109 ($l$ = 79\fdg48, $b$ = +00\fdg15).

\begin{table}[!t]\centering
\setlength{\columnwidth}{0.001\columnwidth}
\setlength{\tabcolsep}{0.8\tabcolsep}
\caption{ Line Fluxes $JH$ and $K_{\rm B}$ Grisms at PA 113$^{\circ}$ }
\begin{tabular}{llccc}
\hline
\multicolumn{2}{c}{} & 
\multicolumn{3}{c}{Regions} \\
\cline{3-4}
\multicolumn{1}{c}{$\lambda$} & 
\multicolumn{1}{c}{Line ID} & 
\multicolumn{1}{c}{~~Ionized Lobes~~} & 
\multicolumn{1}{c}{~~H$_2$ Lobes~~}  \\
\multicolumn{1}{c}{[\AA]} & 
\multicolumn{1}{c}{} & 
\multicolumn{3}{c}{[$\times$10$^{15}$~ergs~~cm$^{-2}$~~s$^{-1}$~~\AA$^{-1}$~]} \\
\hline
10936	&	Pa 6  H~{\sc i}	&	160	&	\dots	&	\\
11621	&	C ~{\sc i} ?  + Fe~{\sc i} ?	&	21	&	\dots	&	\\
11666	&	H$_2$ (3,1) S(4),   He~{\sc i}  ?	&	4.5	&	\dots	&	\\
11892	&	H$_2$ (2,0) S(0)	&	21	&	\dots	&	\\
11970	&	He~{\sc i}	&	4.0	&	\dots	&	\\
12566	&	[Fe~{\sc ii}] ?, + Fe~{\sc ii} ?	&	33	&	1.2	&	\\
12784	&	He~{\sc i}	&	13	&	\dots	&	\\
12817	&	Pa 5 (PaB) H~{\sc i}	&	190	&	11	&	\\
13202	&	?	&	13	&	\dots	&	\\
15335	&	Br 18  H~{\sc i}	&	2.3	&	\dots	&	\\
15549	&	Br 16  H~{\sc i}  	&	1.2	&	\dots	&	\\
15693	&	Br 15  H~{\sc i},  H~{\sc i} 4-15  	&	1.2	&	\dots	&	\\
15875	&	Br 14  H~{\sc i}	&	1.1	&	\dots	&	\\
15995	&	[Fe~{\sc ii}]	&	1.0	&	\dots	&	\\
16102	&	Br 13 H~{\sc i},  H~{\sc i}  4-13  	&	2.5	&	\dots	&	\\
16401	&	Br 12 H~{\sc i},   H~{\sc i}  4-12 	&	2.6	&	\dots	&	\\
16432	&	[Fe II] 	&	15	&	\dots	&	\\
16763	&	?	&	1.7	&	\dots	&	\\
16801	&	Br 11 H~{\sc i},  H~{\sc i} 4-11, Fe ~{\sc i} ? 	&	3.9	&	\dots	&	\\
16997	&	He~{\sc i}	&	2.5	&	\dots	&	\\
17356	&	Br 10  H~{\sc i}	&	5.5    	&	\dots	&	\\
17478	&	H$_2$ (1,0) S(7)	&	1.9    	&	10	&	\\
19570	&	H$_2$ (1,0) S (3) + [Fe~{\sc ii}] ?	&	17	&	61	&	\\
19703	&	H$_2$ (8,6) O (2)	&	1.1    	&	\dots	&	\\
20338	&	H$_2$ (1,0) S(2)	&	3.0    	&	7.7    	&	\\
20580	&	He~{\sc i}	&	4.6    	&	\dots	&	\\
20656	&	H$_2$ (3,2) S(5)	&	0.5	&	\dots	&	\\
20732	&	H$_2$ (2,1) S(3)	&	0.8	&	\dots	&	\\
21133	&	He~{\sc i}	&	0.7	&	1.6    	&	\\
21218	&	H$_2$ (1,0) S (1)	&	6.8    	&	11	&	\\
21658	&	Br 7 (Brg) H~{\sc i}	&	7.4    	&	0.9	&	\\
21892	&	He~{\sc ii}	&	0.9	&	\dots	&	\\
21986	&	?	&	0.7	&	1.0	&	\\
22014	&	H$_2$ (3,2) S(3)	&	1.0	&	\dots	&	\\
22235	&	H$_2$ (1,0) S(0)	&	2.3    	&	4.7    	&	\\
22477	&	H$_2$ (2,1) S(1)	&	1.2    	&	1.7    	&	\\
22870	&	H$_2$ (3,2) S(2)	&	1.0    	&	1.8    	&	\\
\hline
\end{tabular}
\vspace{0.4cm}
\end{table}

\begin{figure*}[!t]
\begin{center}
\includegraphics[width=0.975\columnwidth]{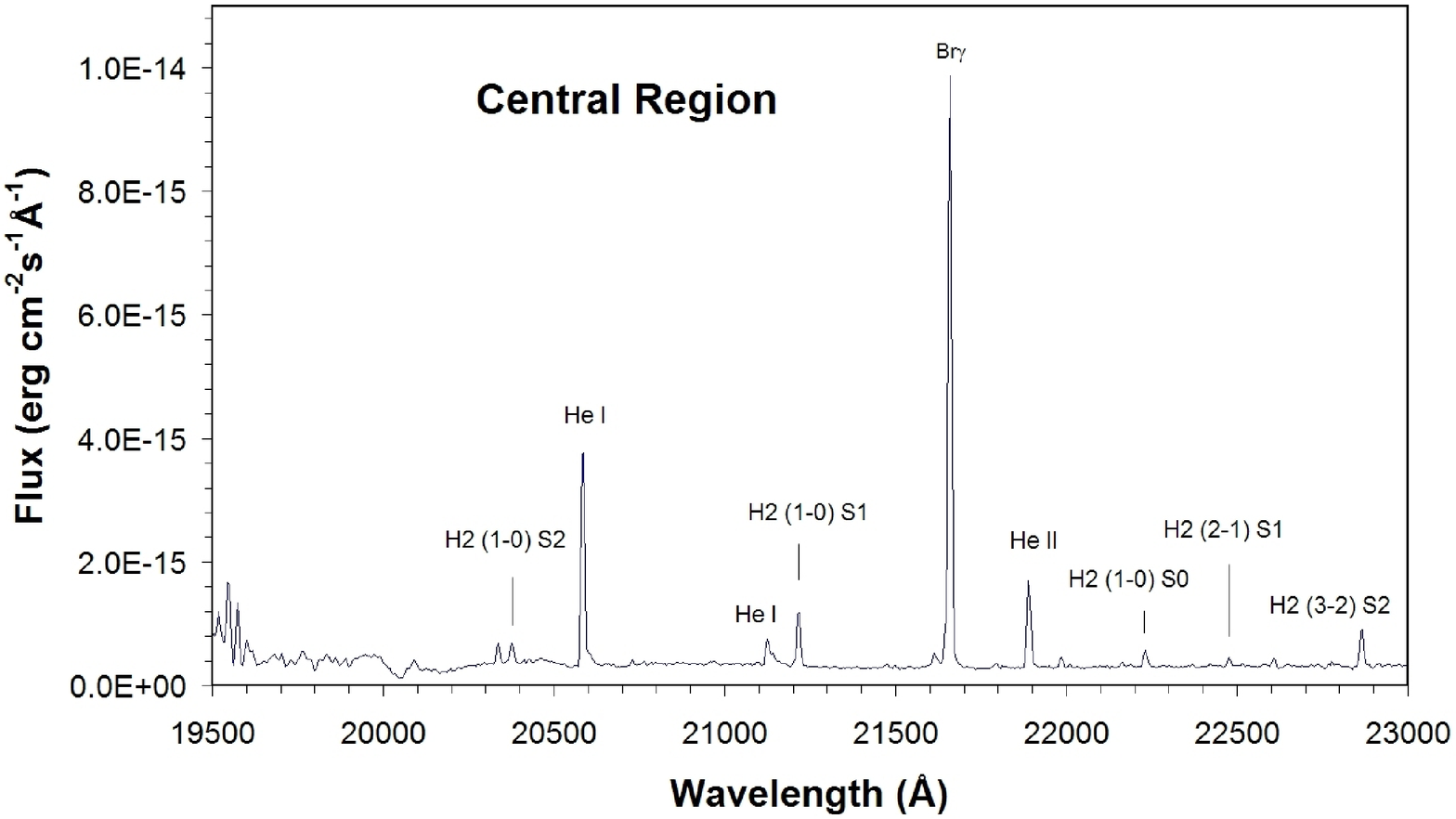}
\hspace*{\columnsep}
\includegraphics[width=0.975\columnwidth]{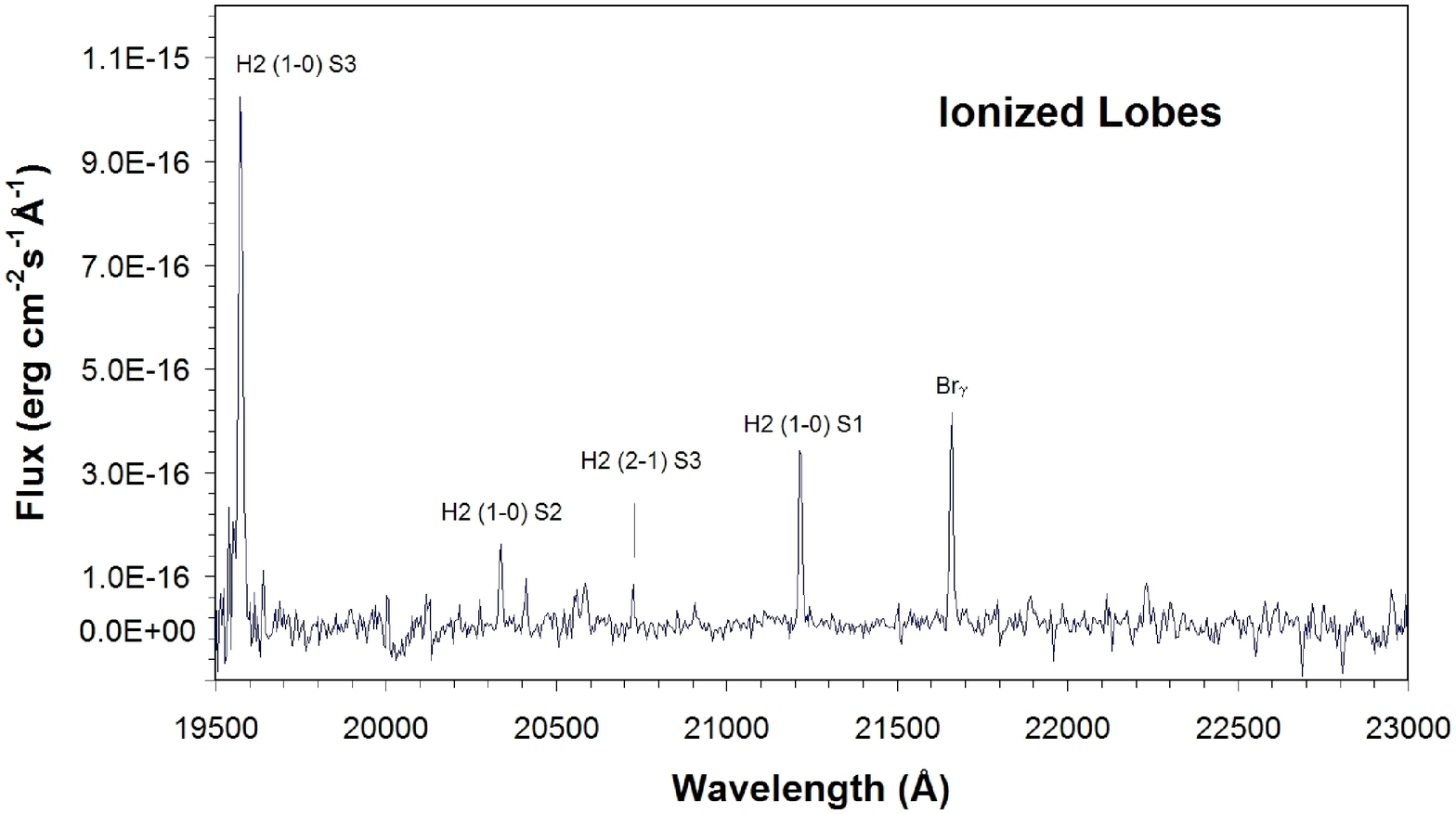}
\vskip .2in
\includegraphics[width=0.975\columnwidth]{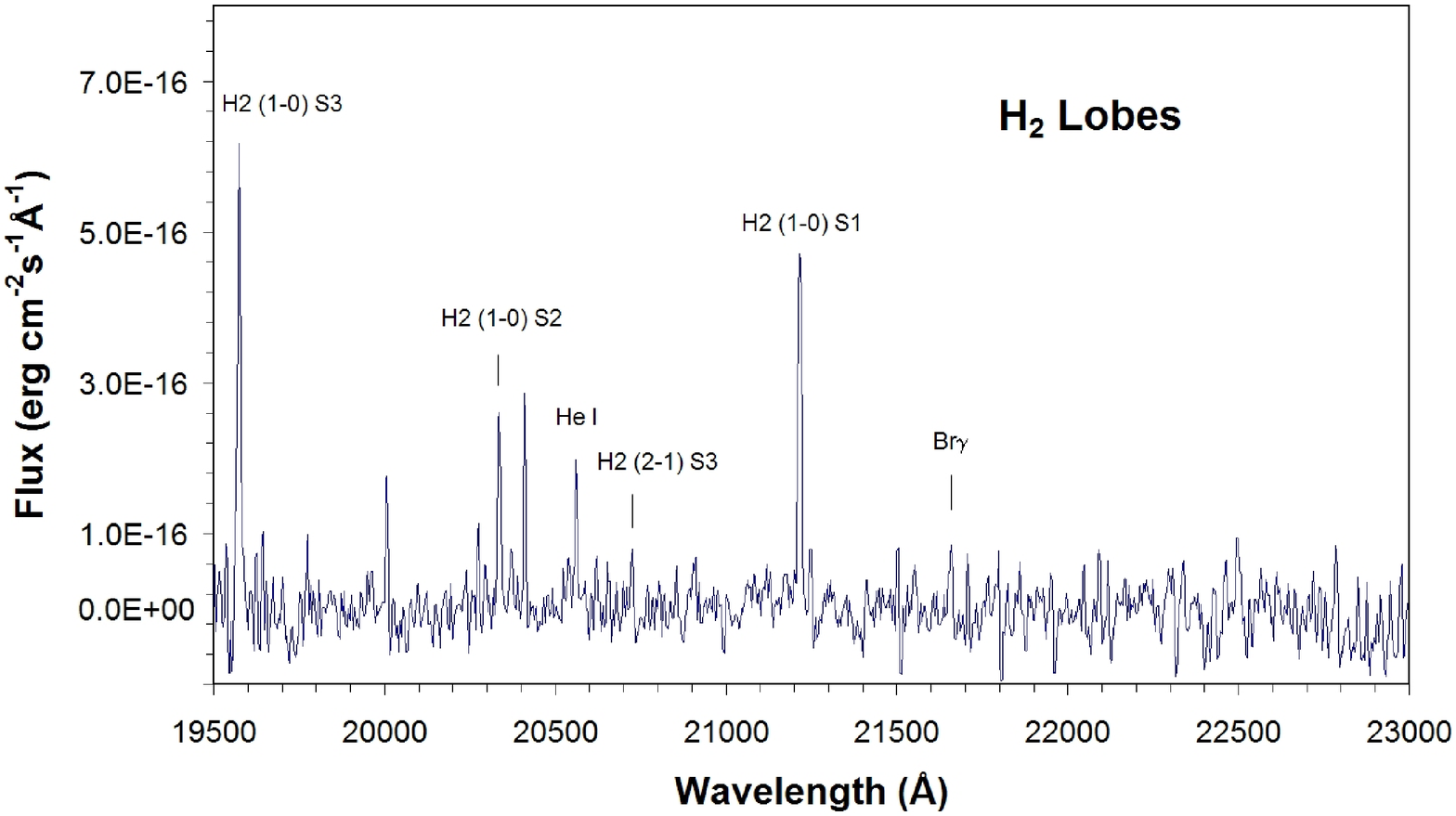}
\hspace*{\columnsep}
\includegraphics[width=0.975\columnwidth]{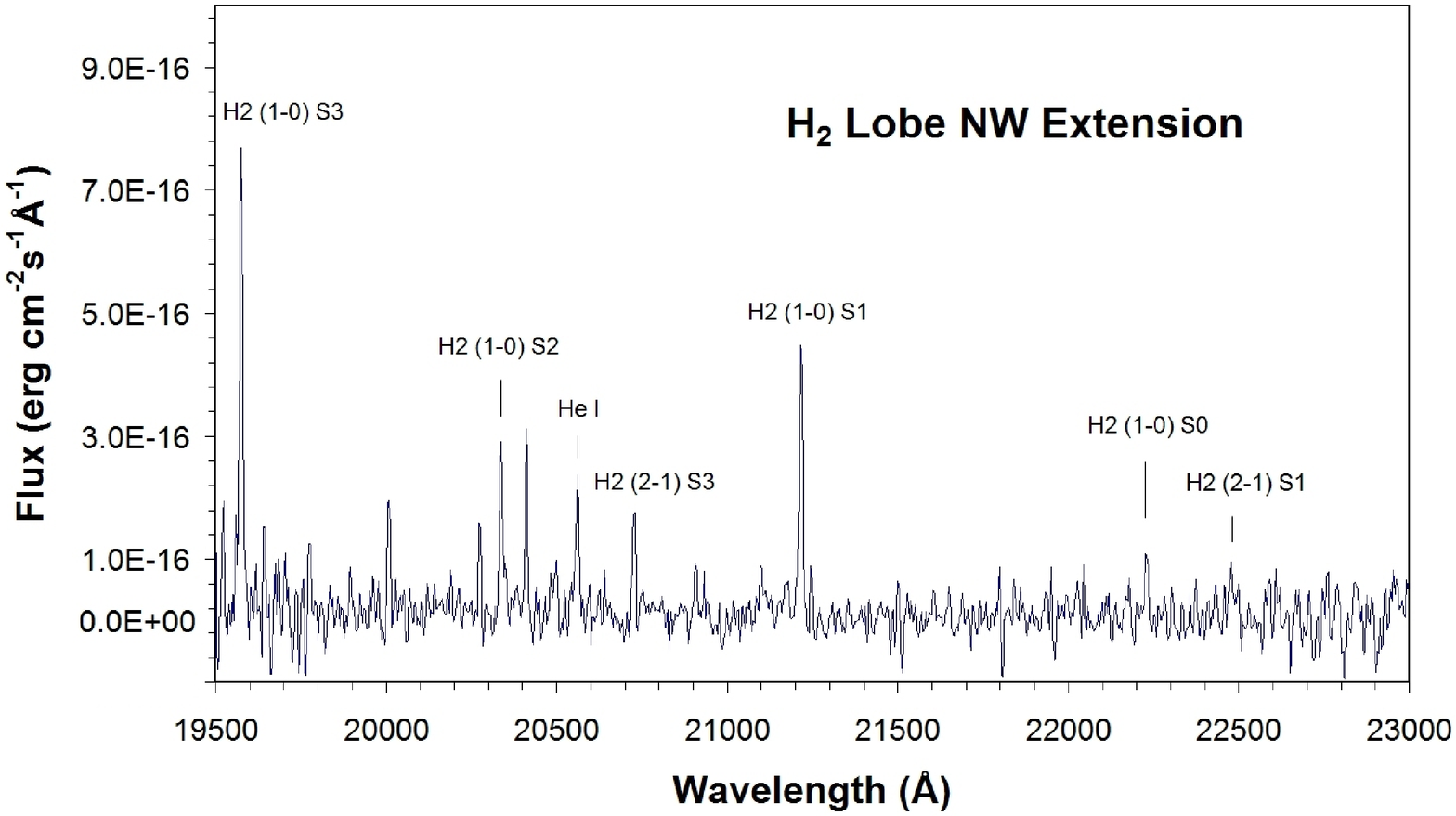}
\caption{
$K_{\rm B}$ spectra along P.A. = 137$^{\circ}$ of the central region, ionized lobes, 
H$_2$-dominated bipolar lobes, and northwest extension of the northwest 
H$_2$ bipolar lobe of NGC\,6881.  
Line identifications are overlaid on the spectra.  
}
\label{fig5}
\end{center}
\end{figure*}

\begin{figure}[!t]
\begin{center}
\includegraphics[width=0.975\columnwidth]{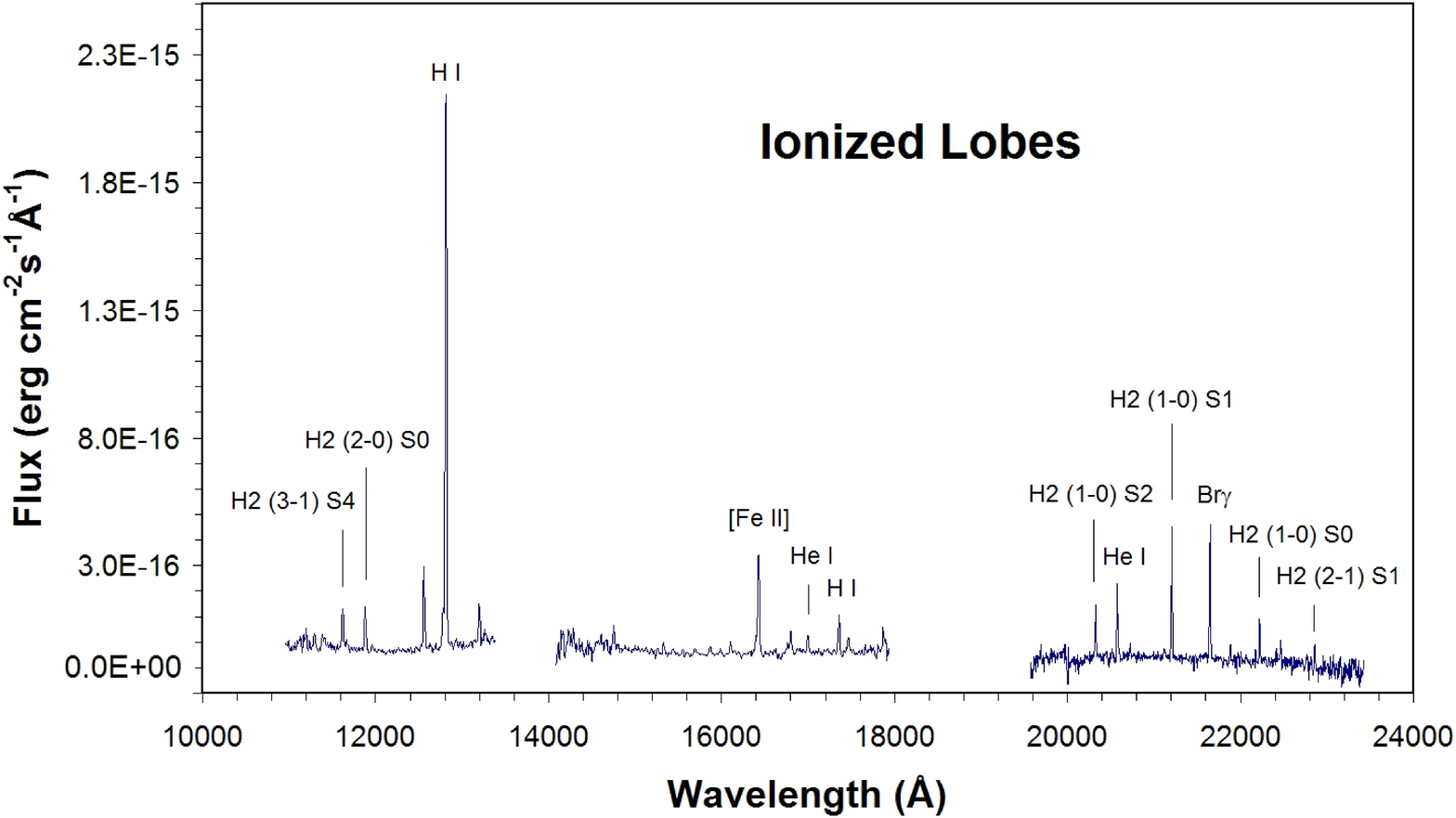}
\caption{
$J$, $H$, and $K_{\rm B}$ spectra along P.A. = 113$^{\circ}$ of the ionized 
lobes of NGC\,6881.  
Line identifications are overlaid on the spectra.  
}
\label{fig6}
\end{center}
\end{figure}

\subsection{Spectroscopic Analysis}

The long-slit spectra of NGC\,6881 detect continuum emission from the 
innermost regions and extended line emission all through the nebula.  
A preliminary inspection of the two-dimensional spectra reveals 
the variation at different locations of the nebula of important 
line ratios (e.g. H$_2$ 1$-$0 S(1)/Br$\gamma$, He~{\sc i}/He~{\sc ii}), 
in agreement with the different spatial distribution of emission shown 
by the Br$\gamma$ and H$_2$ images.  
Guided by these images, we have extracted spectra from four individual 
regions: the central region, the ionized lobes, the H$_2$-dominated 
hourglass lobes, and the northwest H$_2$ lobe extension.  
The $K$ spectra of these regions are shown in Figure~\ref{fig5}, 
and the $JHK$ spectra of the ionized lobes in Figure~\ref{fig6}.
Within our limited spectral resolution, the radial velocity 
derived for the different regions is similar.

The measured line intensities are listed in Table~3 and 4. 
The spectra and line ratios confirm that the central region is 
Br$\gamma$-dominated, the ionized lobes show H$_2$/Br$\gamma$ 
$\sim$ 1, and the outermost regions are H$_2$-dominated.
He~{\sc ii} emission is confined to the central region, while the 
outermost regions show a wealth of H$_2$ lines. 
We note that the spectrum from the ionized lobes is most likely 
contaminated by emission from the H$_2$-dominated hourglass lobes.

\subsection{H$_2$ Excitation}

The molecule of H$_2$ can be excited by shocks or by UV fluorescence.  
The 1--0 S(1)/2--1 S(1) line ratio is traditionally used to diagnose 
the H$_2$ excitation mechanism, with a line ratio $\sim$2 implying UV 
fluorescence \citep{BvD87}, and a line ratio $\sim$10 implying shock 
excitation \citep{Betal89}.  
The 1--0 S(1)/ 2--1 S(1) line ratio in the different regions of NGC\,6881 
ranges from 5.5 up to 7.5, thus indicating that shock excitation can be 
the most likely dominant excitation mechanism.  

\begin{figure}[htbp]
\begin{center}
\includegraphics[width=0.9\linewidth]{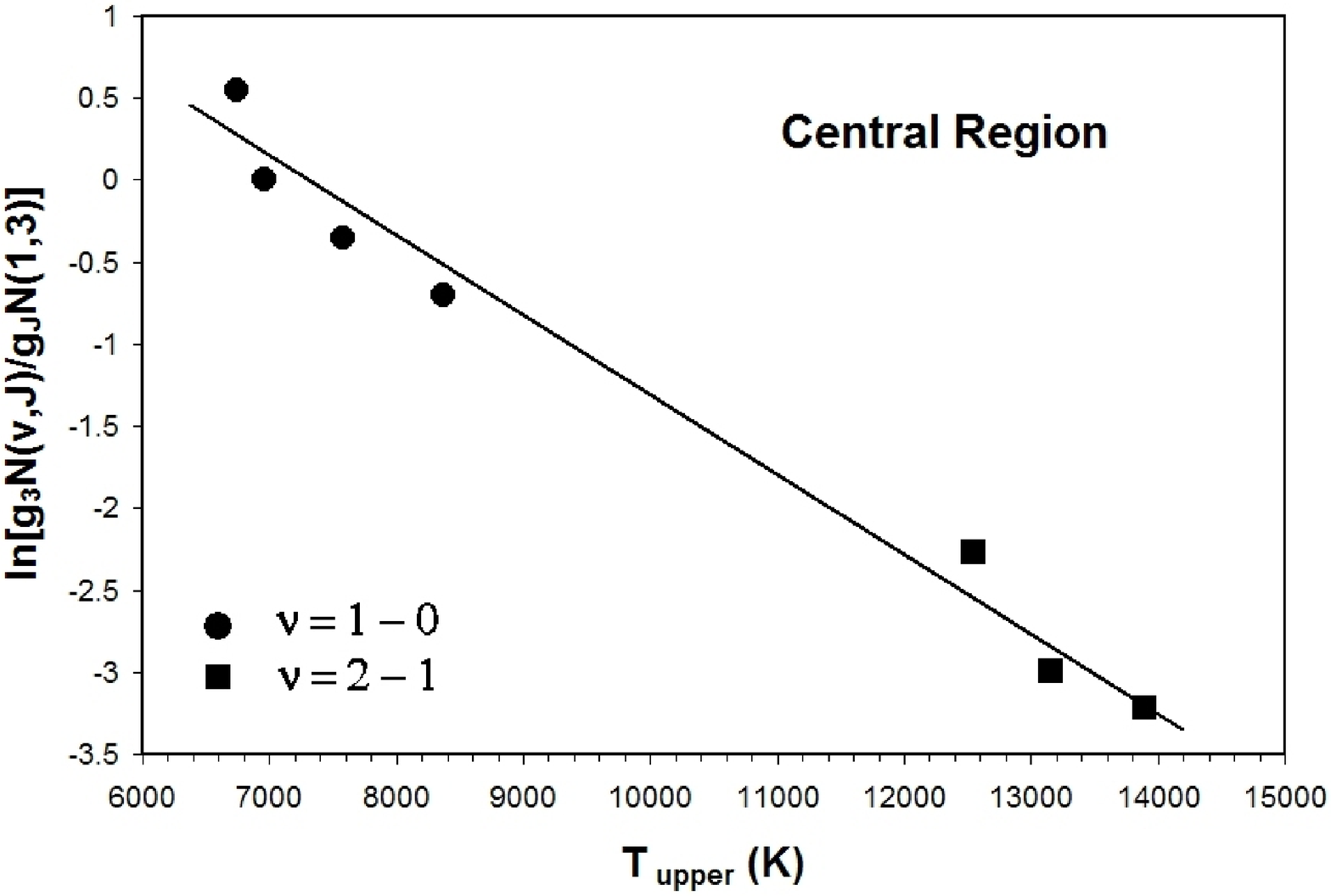}
\includegraphics[width=0.9\linewidth]{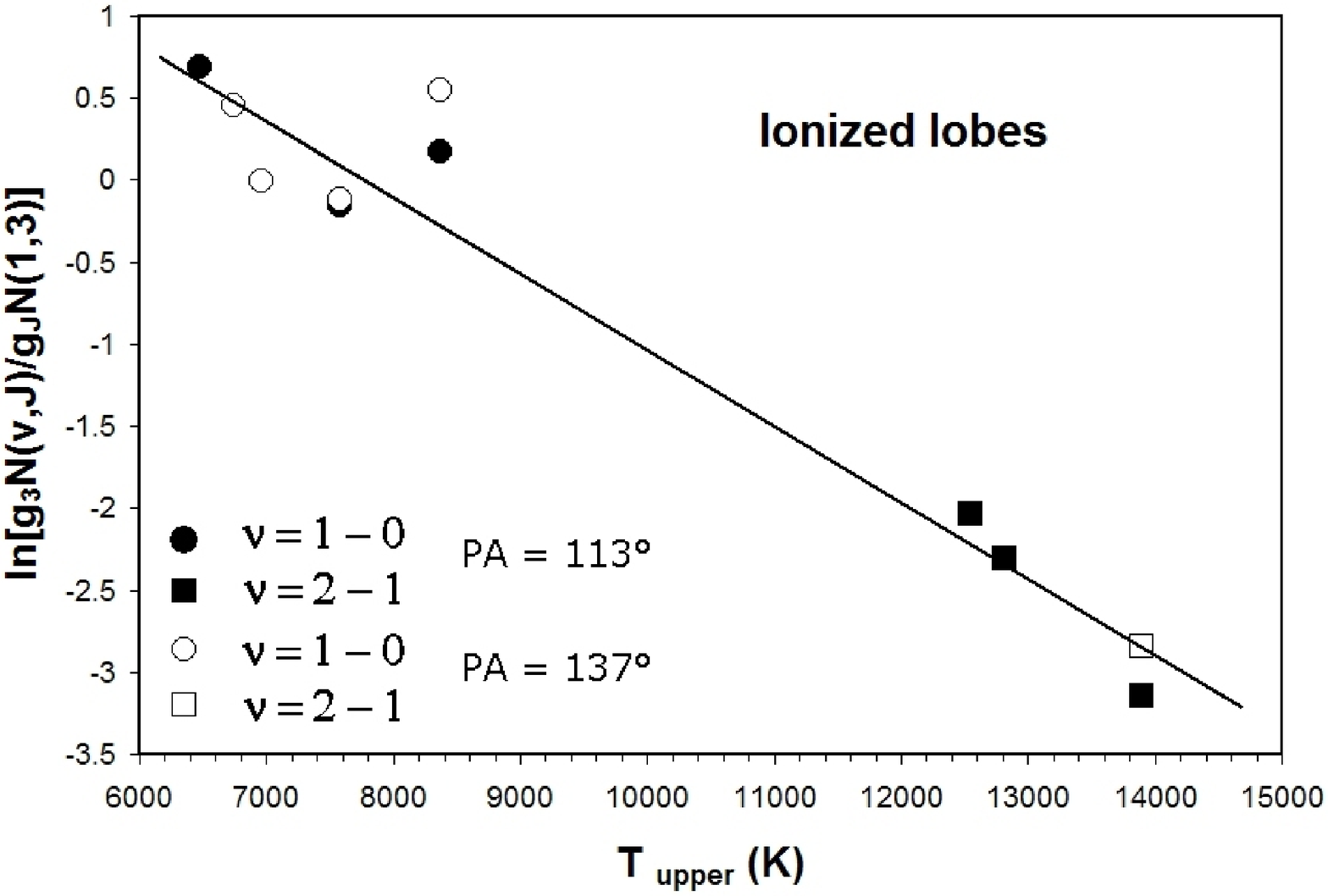}
\includegraphics[width=0.9\linewidth]{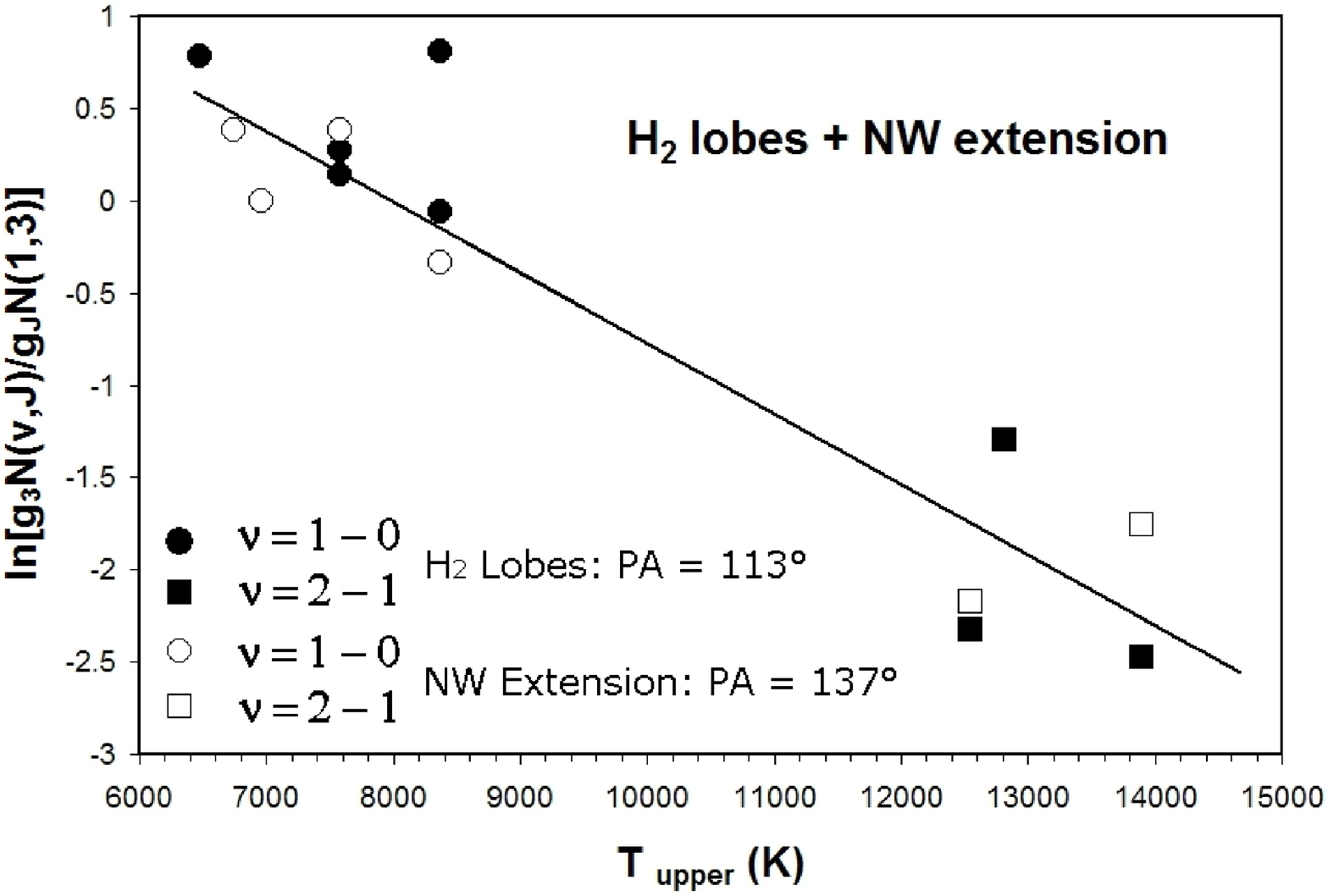}
\caption{
Excitation diagrams for the central region {\it (top)}, ionized lobes 
{\it (center)}, and H$_2$-dominated lobes and northwest extension {\it (bottom)} 
of NGC\,6881.  
The diagrams show the upper state vibration-rotation populations relative 
to the $\nu$\,=\,1, $J$\,=\,3 level plotted against the upper state energy.  
The linear fits to the populations of different vibrational levels find 
rotational excitation temperatures of 2000~K, 2100~K, and 2700~K for 
the central region, ionized lobes, and H$_2$-dominated lobes and northwest 
extension, respectively. 
} 
\label{fig7}
\end{center}
\end{figure}

In a case of pure collisional excitation, the population of the energy 
levels of the H$_2$ molecule are described by a single-temperature 
Boltzmann distribution.  
To verify this statement, the excitation diagrams for several regions 
of NGC\,6881 (Figure~\ref{fig7}) have been calculated using a Boltzmann 
distribution of states where the column densities are related to the 
$\nu$\,=\,1, $J$\,=\,3 state by 
\begin{equation}
\label{b1}
\frac{g_3N(\nu,J)}{g_{\rm J}N(1,3)} = 
\exp \left (-\frac{E(\nu,J)-E(1,3)}{kT_{ex}}\right),
\end{equation}
$g_{\rm J}$ and $g_3$ being the statistical weights. 
This equation can be rearranged in terms of the flux
\begin{equation}
\label{b2}
\frac{F(\nu',J')\nu_{1,0 S(1)}A_{1,0 S(1)}g_3}{F(1,3)\nu_{\Delta\nu,\Delta J}A_{\nu',J'\rightarrow\nu'',J''}g_{\rm J}} = 
\exp \left (-\frac{E(\nu,J)-E(1,3)}{kT_{ex}}\right), 
\end{equation}
where $A_{\nu',J'\rightarrow\nu'',J''}$ represent the transition 
probability, and $F(\nu',J')$ is the observed flux at frequency 
$\nu_{\Delta\nu,\Delta J}$.  
  
The temperature can be inferred from the slope of the line fitted to the 
data points in Fig.~\ref{fig7}.  
In NGC\,6881, the value of the vibrational excitation temperature, 
$T_{ex}$($\nu$), is similar to this of the rotational excitation 
temperature, $T_{ex}$($J$), and thus it can be concluded that shocks 
are the dominant excitation mechanism of H$_2$.  
The excitation temperature is 2100 K for the central region, 
2200$\pm$100~K for the ionized and H$_2$ lobes, and 2700$\pm$100~K 
for the extension of the NW H$_2$ lobe.  
There is thus, a trend for the temperature to increase as we move 
farther from the central regions of NGC\,6881.

\section{Discussion}

\subsection{Ionized and Molecular Gas in NGC\,6881}

The H$_2$ emission of bipolar PNe is typically found in their equatorial 
rings and on the walls of their bipolar lobes, outlining closely the 
distribution of ionized material shown in H$\alpha$ or [N~{\sc ii}] 
images \citep{Ketal96,Getal00}.  
In NGC\,6881, however, the different spatial distributions of ionized 
material and molecular hydrogen trace distinct nebular structures.  
A search in the literature shows that there are very few cases 
of bipolar PNe and proto-PNe in which the H$_2$ and H$\alpha$ 
or [N~{\sc ii}] morphologies are different:  CRL\,2688, M\,2-9, 
NGC\,2440, NGC\,7027, J\,900, and Hb\,12. 
From this short list, NGC\,2440 has to be discarded as recent 
\emph{HST} images (Proposal ID.\ 11090) have shown that the 
H$_2$ arc seen by \citet{Letal95} follows the outermost 
H$\alpha$ emission.  
In CRL\,2688 (the Egg Nebula), the H$_2$ emission is distributed 
along the equatorial plane and at the tips of the bipolar lobes 
seen in scattered light \citep{Letal93,Setal98}.  
The H$_2$ bipolar lobes of M\,2-9 envelop the bipolar lobes seen in 
emission lines of ionized species such as Fe~{\sc ii} \citep{SBG05}.  
The four-lobed shell of H$_2$ in NGC\,7072 surrounds its elliptical 
ionized core, revealing the location of a photodissociation region 
(PDR) that is excited by the absorption of UV photons 
\citep{Hetal99,Letal00}.  
In J\,900, the H$_2$ emission shows a linear structure 
extending $\sim$40\arcsec\ that ends in an arc-like 
structure \citep{Setal95,Letal95}.  
This feature has been suggested to be caused by UV excitation 
of stellar radiation that leaks out through holes in the 
nebular envelope \citep{Setal95}.  
Finally, the inner ionized bipolar lobes of Hb\,12 are surrounded by an 
H$_2$ eye-shaped structure that is interpreted as a broad equatorial 
ring and a larger pair of bipolar lobes \citep{HL96,Wetal99,KH07}.

Therefore, among the ``exclusive'' group of PNe with different distributions 
of H$_2$ and ionized gas, M\,2-9, Hb\,12, and NGC\,6881 have ionized bipolar 
lobes that are enclosed within H$_2$ bipolar lobes.  
We note, however, that Hb\,12 is a notable case of UV-excited fluorescent 
H$_2$ emission \citep{LR96,HL96} and the same is suspected for M\,2-9 
\citep{SBG05}, while NGC\,6881 is representative of collisionally excited 
H$_2$ emission.  
The presence of multiple pairs of bipolar lobes in PNe is further discussed 
in the following section.  

\subsection{Multiple Bipolar Ejections in NGC\,6881}

The occurrence of multiple ejections of material is common among PNe, 
particularly among type I multiple-shell PNe that show faint, giant 
haloes \citep{Cetal87}.  
The haloes, interpreted as the relic of late thermal-pulse episodes 
\citep{SP95,Hetal97}, are mainly associated with elliptical or round 
PNe, but are rare among bipolar PNe.  
Recurrent bipolar ejections resulting in multiple independent pairs 
of bipolar lobes have been reported for only a handful of PNe.  
The different pairs of bipolar lobes may have similar morphology and 
orientation, e.g., M\,2-9 \citep{B99,Setal97}, but in most cases they 
show notable differences, either affecting their morphology and degree 
of collimation, e.g., Mz\,3 \citep{Getal04,SGetal04}, the orientation 
of their symmetry axes, e.g., M\,2-46 \citep{Metal96}, or its 
ionization degree, e.g., Hb\,12 \citep{Wetal99}.  
In NGC\,6881, we find two pairs of highly collimated ionized lobes 
oriented along slightly different directions and a pair of H$_2$ 
bipolar lobes with an open hourglass morphology.

The two pairs of ionized lobes of NGC\,6881, first reported by 
\citet{Metal96}, have been described in detail by \citet{GM98} 
and \citet{KS05}.
Precession is evidenced by the different symmetry axes of each pair 
of ionized lobes and from the alignment between the equatorial ring 
and the youngest pair of bipolar lobes.  
The sharp morphology, similar kinematical age, and different orientation 
of the two pairs of bipolar lobes point to a sudden collimated ejection 
of material that has carved out bipolar cavities into the surrounding 
medium.  
The morphology of the ionized bipolar lobes of NGC\,6881 is very 
reminiscent of the proto-PN CRL\,618 in which high-velocity jets 
moving along different directions \citep{Cetal03} have produced 
bipolar cavities in the nebular envelope \citep{TG02}.  
The similarities between the two bipolar nebulae extend to the coaxial 
rings of the bipolar cavities and to the bright, compact knots at their 
tips.  
The collimated outflows of CRL\,618 might well evolved in the future 
into bipolar lobes similar to those of NGC\,6881.

The limb-brightened morphology of the H$_2$ lobes indicates that the 
material is mainly confined in their walls, thus suggesting that the 
bipolar lobes carry enough momentum to sweep up and press the 
circumstellar medium into a thin sheet.  
Indeed, the prevalent shock excitation of H$_2$ requires a source 
of momentum.  
It is unlikely that a current fast ($\sim$1000~km~s$^{-1}$) wind from 
the central star would provide this momentum, as such a stellar wind 
would be trapped by the inner ionized lobes.   
Our spectra do not provide information on the expansion velocity of 
the H$_2$ lobes, but an upper limit can be derived assuming that the 
[N~{\sc ii}] emission from the loop-like structure at the edge of the 
southeast lobe is produced by a shock.  
For a planar shock to be able to produce significant [N~{\sc ii}] 
$\lambda$6584~\AA\ emission, but low or minimal [O~{\sc iii}] 
$\lambda$5007~\AA\ emission \citep{GM98}, the shock velocity must 
not exceed $\sim$60~km~s$^{-1}$ \citep{HRH87}.  
Such high speed makes unlikely the origin of the momentum in the thermal 
pressure of hot gas within the bipolar lobes.  
However, the momentum of the H$_2$ bipolar lobes must have been provided 
at the ejection time, either in a sudden bipolar ejection or through a 
stellar wind intrinsically bipolar or isotropic but collimated by an 
equatorial structure.  
The oblique shock of a bipolar lobe expanding at $\sim$60~km~s$^{-1}$ 
would be able to shock excite the H$_2$ molecules along its edge, 
while giving rise to [N~{\sc ii}] and H$\alpha$ emission at its 
tip.  
Such ballistic ejection may also explain the increasing H$_2$ temperature 
excitation with the distance to the nebula center.

The very different collimation degree of the ionized and H$_2$ lobes of 
NGC\,6881 suggests that they formed during two different ejection events 
with substantial changes in the collimation conditions or even in the 
collimation mechanisms between the two episodes.  
The H$_2$ lobes formed very likely by the action of an episodic wind 
producing an hourglass-shaped ejection, while high velocity jets 
ejected along close directions produced the two pairs of ionized lobes.  
An additional period of isotropic heavy mass loss resulting in a 
dense envelope is revealed by concentric arcs detected around the 
innermost regions of NGC\,6881 \citep{Cetal04}.

\subsection{NGC\,6881 and the Surrounding Interstellar Medium}

The morphology of the southeast and northwest H$_2$-dominated bipolar 
lobes are discrepant, with the former being smaller than the latter 
and showing a sharp edge at its southeast tip.  
This sharp edge is spatially coincident with a [N~{\sc ii}] loop-like 
feature (Fig.~\ref{fig4}), a structure that has been interpreted as a 
low velocity precessing jet \citep{GM98}, but that, in view of the 
perfect spatial coincidence with the southeast tip of the H$_2$ lobe, 
seems more likely to mark its termination.  
The abrupt border of the southeast H$_2$ lobe and the [N~{\sc ii}] 
loop-like structure may unveil the interaction of this bipolar lobe 
lobe with dense material.

As we noted in $\S$3.1, NGC\,6881 is embedded within diffuse emission 
that forms part of Sh\,2-109, an inhomogeneous H~{\sc ii} region located 
at a distance of 1.4$\pm$0.4 kpc and with a diameter of 400$\pm$100 pc 
\citep{FB84}.  
In view of the hint of interaction of the southeast H$_2$ lobe of 
NGC\,6881 with dense material, resulting in the notably different 
morphologies of the southeast and northwest lobes, it is worthwhile to 
investigate the physical association of NGC\,6881 with Sh\,2-109.  
The distance to NGC\,6881 has been estimated to be 2.5 kpc by \citet{CKS92} 
who used the Shklovsky method according to the Daub scheme.  
In their work, Cahn et al.\ used an angular radius of 2\farcs5 
which seems inadequate as it only traces the bright, central 
regions of NGC\,6881.  
This region has an angular radius $\sim$1\farcs2, but if we account for 
the ionized bipolar lobes, then an averaged angular radius of 6\farcs5 
has to be used.  
If we adopt this radius, then the revised distance to NGC\,6881 
using the Cahn et al.\ method would be $\sim$1.5 kpc, supporting its 
physical association with the H~{\sc ii} region Sh\,2-109.

An inspection of IPHAS H$\alpha$ images of this region reveals that 
the diffuse emission in which NGC\,6881 is embedded constitutes the 
eastern wall of a large cavity-like structure.  
H~{\sc i} 21~cm line observations of the neutral gas distribution 
and kinematics of the region confirm that this structure is a cavity 
with an expansion velocity $\sim$10~km~s$^{-1}$ \citep{Cetal96}.  
The radial velocity of this H~{\sc i} cavity, 
$v_{\rm LSR}\sim+16$~km~s$^{-1}$, is very similar to the radial velocity 
of NGC\,6881, $v_{\rm LSR}\sim+2$~km~s$^{-1}$ \citep{GM98}.  
The projection of the nebula onto the H~{\sc i} cavity and their 
similar radial velocities suggest that NGC\,6881 may be located 
within this cavity, with its H$_2$-dominated southeast lobe hitting 
onto the dense material at the wall of the cavity, slowing it down 
and producing a sharp edge, while the northwest H$_2$ lobe expands 
into the H~{\sc i} cavity with little opposition.

\section{Conclusions and Summary}

New images and near-IR spectra of NGC\,6881 confirm the different 
distribution of ionized material and molecular hydrogen within this 
PN.  
The H$_2$ images have resolved its central region which shows the spatial 
distribution expected for a ring of molecular material surrounding an 
innermost ionized ring.  
The H$_2$ bipolar lobes share the same orientation than the ionized 
lobes, but they are less collimated and display a distinct hourglass 
morphology.  
The H$_2$ emission is predominantly shock excited, both in the 
central ring and in the H$_2$ bipolar lobes.

We infer a complex history formation for NGC\,6881, involving important 
changes in the collimation mechanism of the bipolar lobes in the latest 
evolutionary stages.  
The H$_2$ bipolar lobes are consistent with the action of a bipolar 
wind.  
The final evolution of the PN implied the ejection of episodic fast 
collimated outflows with changing directions that interacted with 
the dense circumstellar envelope ejected in a previous phase.

The southeast H$_2$ lobe of NGC\,6881 is less extended than the northwest 
lobe and shows a sharp edge, implying the interaction of NGC\,6881 with an 
inhomogeneous interstellar medium.  
We find very likely that NGC\,6881 is within a cavity in the midst of 
the H~{\sc ii} region Sh\,2-109.

\acknowledgments

This work is funded by grant PNAYA2005-01495 of the Spanish MEC.
G.R.L. thanks the IAA for its hospitality and the support of a postgraduate 
scholarship from CONACyT (Mexico).   
We thank J.\ Acosta and G.\ G\'omez for taking the LIRIS images of 
NGC\,6881, 
and K.\ Viironen and the IPHAS team for providing us with large-scale 
H$\alpha$ images of the region around NGC\,6881.  
We are also grateful to E.J.\ Alfaro for enlightening discussion on the 
location of NGC\,6881 within the Galaxy.



\clearpage

\clearpage


\end{document}